# Closed ASL Interpreting for Online Videos

Raja Kushalnagar
Information Technology Program
Gallaudet University
raja.kushalnagar@gallaudet.edu

Matthew Seita
Computer Science Department
Rochester Institute of Technology
mss4296@rit.edu

Abraham Glasser
Computer Science Department
Rochester Institute of Technology
atg2036@rit.edu

## ABSTRACT
Deaf individuals face great challenges in today's society. It can be very difficult to be able to understand different forms of media without a sense of hearing. Many videos and movies found online today are not captioned, and even fewer have a supporting video with an interpreter. Also, even with a supporting interpreter video provided, information is still lost due to the inability to look at both the video and the interpreter simultaneously. To alleviate this issue, we came up with a tool called closed interpreting. Similar to closed captioning, it will be displayed with an online video and can be toggled on and off. However, the closed interpreter is also user-adjustable. Settings, such as interpreter size, transparency, and location, can be adjusted. Our goal with this study is to find out what deaf and hard of hearing viewers like about videos that come with interpreters, and whether the adjustability is beneficial.

## 1. Introduction
It can be difficult for people with sensory disabilities, to obtain equal access to information. Deaf and hard of hearing viewers usually need visual access to aural information in online videos as they cannot understand the aural information. The majority of videos online are not captioned and even fewer show sign language translation via an interpreter. For many deaf viewers, captioning can be challenging to follow, because the speed of verbatim captioning is likely to exceed their reading abilities [2]. Even after controlling for reading level, deaf students still learned less from on-screen text than hearing peers, apparently because of differences in background knowledge and information processing strategies [6]. Partly because they find captions hard to follow and partly because they get less information, they prefer to ASL interpreters over closed captions. Videos that are interpreted still have issues that must be addressed. While there is little research on how deaf people perceive a recording of a sign language interpreter when watching online videos, some problems are obvious.

For example, some loss of information and understanding is inevitable with a deaf person watching an interpreted video due to having to shift the eyes side to side between two locations on the screen. In doing so, if the eyes are on the interpreter, the viewer could miss information on the video, or vice versa. Another issue that could arise is that different viewers have different preferences regarding the interpreter; perhaps some prefer different interpreter video sizes or some prefer the interpreter video in different locations on the screen. In conducting this study, we attempted to learn more about how deaf people perceive interpreted videos and what their general preferences are.

The goal of the research presented in this paper is to assess what features and qualities of closed interpreting appeals to deaf viewers. In this paper, we propose the idea of "closed interpreting" which takes the idea of closed captioning and applies it to interpreting. While the idea is similar to closed captioning, closed interpreting as described in this paper is much more dynamic, allowing viewers to adjust the interpreter as they please. To accomplish this, we developed a tool that allows the closed interpreter to be manually adjusted by the viewers to adhere to their preferences. For example, the viewer may change the size, location, or transparency of the interpreter. Additional features may be toggled on or off, such as the ability to pause or resume the interpreter or the ability to slow down and speed up the interpreter. Additional research was also done with eye tracking software to automatically pause the interpreter when the viewer is not looking at them and is looking at the contents of the video instead. Once the viewer resumes their gaze on the interpreter, the interpreter resumes and speeds up to catch up to the current point in the video. To apply this tool, online videos will presumably come with the option to turn on or off closed interpreting, just as with closed captioning, although the maker of the video will still have to be willing to provide the interpreter and make the recording, and upload it with their video. This would allow deaf people to turn on closed interpreting as necessary, and the creator of the videos does not have to make two versions of the same video, one with interpreting and one without.

For evaluation, we conducted a user study. To analyze the efficiency of the tool and to discern user preferences, we recruited deaf and hard of hearing volunteers to test our software. They were told to perform a sequence of steps, and they also watched a video with closed interpreting available, and were allowed to fiddle with some of the features. Finally, they answered a series of questions designed to find out which features were most desired, and whether or not the tool was helpful towards their understanding of the video contents, using questions with a Likert scale and open-ended questions. The research presented in this paper could be helpful towards future work done on closed interpreting by deducing which features are most important and thus should be focused on, and what users liked the most about this idea.

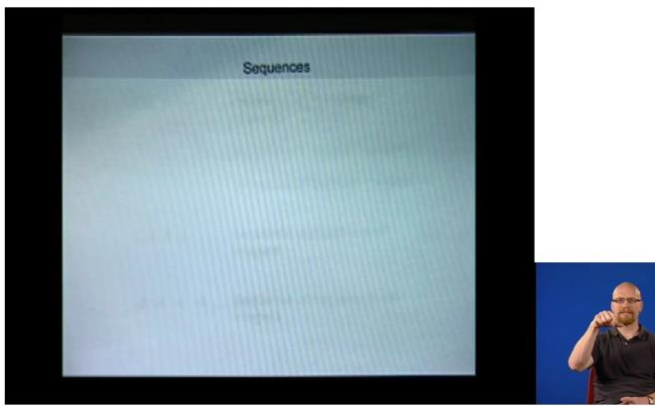

**Figure 1: Example of what "static" closed ASL interpreting looks like.**

## 2. Related Work

Prior research work has shown that deaf and hard of hearing viewers do not get equal access to videos, compared with their hearing peers. In general, hearing viewers are able to listen to the verbal information and attend to the visual information simultaneously [7]. On the other hand, deaf viewers usually cannot view the visual translation of auditory information simultaneously with the video visual information. For example, in classrooms with sign language interpreters, deaf students have to focus on the interpreter or the classroom visuals at different times. This focus challenge becomes harder with a third source of visual information (e.g., slides or computer screens), as they risk losing the thread of a lecture because the different information sources will be out of synchrony and they likely will be unable to predict which source is more important at any given time [6]. Deaf students learn less than their hearing peers research on accessible views for interpreters focused on the ability for deaf students to replay captioning if they missed information, and whether or not this feature was helpful [5]. Research with visual cues in the classroom has also been done [3]. This paper analyzed the benefits of having visual cues during lectures, so that deaf students know when to change gaze from the interpreter to the slides (such as when a new slide appears or when the slide is referenced by the teacher) or other modes of information. An interface for YouTube videos has also been created that allows for pausing and resuming of captions to allow more time for absorbing information, and the effectiveness and helpfulness of this interface was evaluated [4]. Other work has been done on finding out the maximum replay speed that deaf people can accurately understand sign language interpreters. This value was found to be approximately 2 times the normal speed of signing [5]. Additionally, Cavender et al. gives insight on which configuration of the screen was most preferred by deaf and hard of hearing viewers, which could be helpful in finding out how to best arrange the video and interpreter on the screen for this study [1].

Overall, part of the research done in this paper builds upon several previous works done on this topic, and incorporates these ideas into the tool that we developed. However, some of the ideas in this study (e.g. finding out the best size and location of the interpreter video on the screen) have not had much prior research.

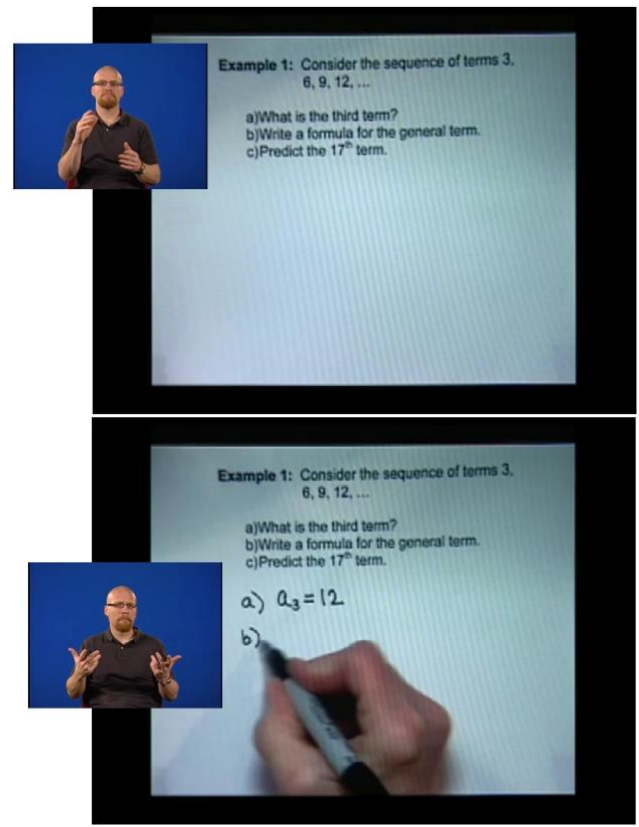

**Figure 2: Example of what "tracked" closed ASL interpreting looks like.**

## 3. Methodology

To implement closed interpreting, we used a lecture video that taught mathematics sequences. We also obtained an interpreter video that corresponds to the lecture video and both videos were embedded in an HTML browser. In the static closed interpreting implementation, both the interpreter and lecture videos are side by side and immovable. An example of what it looks like is shown in Figure 1.

Figure 2 shows the idea behind the tracked interpreter implementation. In tracked interpreting, the interpreter video follows the contents of the lecture video such that the interpreter video is lined up horizontally with the relevant area of interest in the lecture video (e.g. when the instructor is writing something, the interpreter is lined up horizontally with the writing). The tracking was done manually a human person looked at the lecture video and decided the best place for the interpreter to be in each particular moment. In Figure 2, the interpreter starts in line with the notes in the lecture video as shown in the top image, then when the instructor starts writing in the empty space in the middle, the interpreter jumps to the position shown in the bottom image, "tracking" the lecture materials.

Finally, in the customizable closed interpreting, the interpreter video is allowed to be adjustable, but not the lecture video. The position, size, and transparency of the interpreter video can all be adjusted. The abilities to pause and play the videos were also included. The interpreter video also has hide and show functionality.

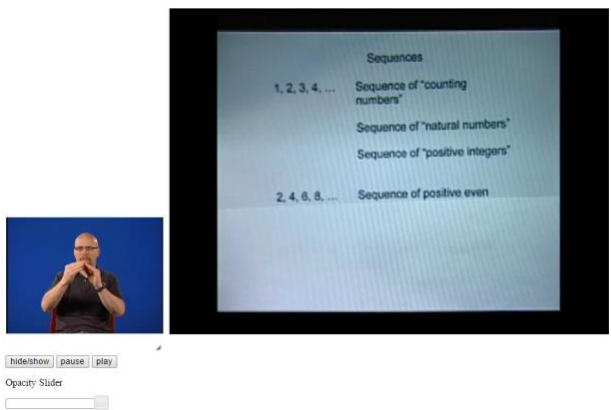

**Figure 3: Example of "Customizable" closed interpreting.**

Figure 3 shows the layout we used for the customizable closed interpreting. The interpreter screen is allowed to be moved by clicking and dragging, but the lecture video is static. The interpreter video can be resized using the small black arrow in the lower right corner of the video and clicking and dragging it. There is also a slider labeled "Opacity" that allows the transparency of the interpreter video to be adjusted from 10.

## 3.1 Experiment and Survey Implementation

To conduct an evaluation of our tool, we recruited 19 deaf and hard of hearing survey participants. This criterion was established because the project evaluates how well deaf and hard of hearing viewers can understand the video with the closed interpreting, as well as their preferences. To do that, the participants must have hearing loss and must have a good understanding of sign language. The experiments and surveys took a half hour to complete, and the participants were reimbursed ten dollars for their time.

The survey consisted of a pre-experiment section, used to determine eligibility for the study as well as demographic information. If any participant was not eligible, they were promptly dismissed from the study. The remaining participants were then walked through the experiment. The experiment was structured so that participants watched three YouTube lecture videos with closed interpreting. One of the videos included "static" closed interpreting, as shown in Figure 1, where the lecture video and the interpreter video were simply set side by side and were non-adjustable. The second video included "tracked" interpreting, where the interpreter video follows the contents of the lecture video as described in section 3.1. The other video included "customizable" interpreting, where the lecture video was still static.

The participants watched videos in different order to reduce the bias that comes with watching a certain video, first or watching a certain video before or after another one. After watching the videos, participants filled out the post-experiment survey that asked them to rate their understanding and experience with each of the two videos. For each of the videos, they were asked to rate, using a Likert scale of 1 to 5, their satisfaction with the closed interpreter as well as their ability to see and understand the contents of the video. For the customizable interpreting video only, they were asked to rate, again using a Likert scale of 1 to 5, how much they liked the ability to move, change the size of, and change the transparency of the interpreter. They were asked to explain each of their responses if they could. They were also asked open-ended questions relating to their preferences.

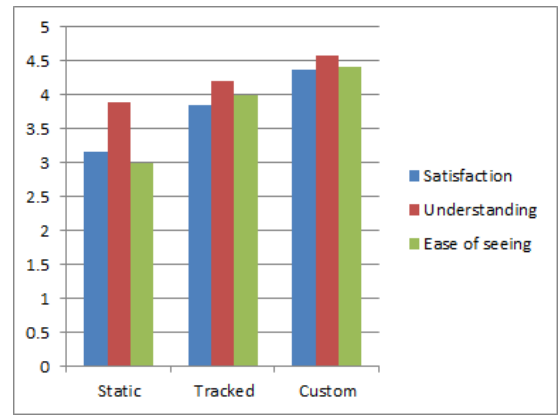

**Figure 4:** Average user responses (on a scale of 1 to 5) for how they felt about each of the three interpreter implementations

## 4. Experimental Results

After conducting the experiments, we collected the Likert scale results, answers to open-ended questions, and other feedback. Each of the three implementations had three Likert scale questions in common, relating to user satisfaction, understanding of the lecture video contents, and how easy it was to view the lecture and interpreter videos. The Likert scale had a range of 1 to 5, with 1 being the lowest rating and 5 being the highest rating. The mean for satisfaction, understanding, and ease of viewing for the static interpreting video was 3.16, 3.89, and 3, respectively. The mean for satisfaction, understanding, and ease of viewing for the tracked interpreting video was 3.84, 4.21, and 4, respectively. The satisfaction, understanding, and ease of viewing for the customizable interpreting video was 4.36, 4.58, and 4.42, respectively, as shown in Figure 4. The survey section for the customizable interpreting implementation also had three additional Likert scale questions. The questions asked how they felt about each of the three features, being able to move, change the transparency of, and resize the interpreter. They were scaled the same as described in the previous paragraph, from 1 to 5, with 1 being the lowest rating (not helpful) and 5 being the highest rating (very helpful). The mean value of the responses for the ability to move, change transparency, and resize were 4.42, 3.42, and 4.31, respectively as shown in Figure 5.

To analyze the results for statistical significance, we used the Mann-Whitney U test. The data for which we tested the significance was derived from the three Likert scale questions that all 3 implementations had in common, i.e., the questions relating with satisfaction of each implementation, how well the viewer understood the lecture for each implementation, and how easy it was to see the lecture information for each implementation. Figure 6 showcases this data.

There were 3 questions and 3 implementations, and the results for each question was compared across the other two implementations, there were 9 total comparisons. Sample A and B refer to the two specific questions being compared. In these two columns, the letter S refers to the question regarding satisfaction, the letter U to understanding, and the letter E to ease of viewing information.

**Table 1: Comparison of static, tracked and custom implementations**

| Sample A | Sample B | $U_A$ | $S_{.05}$ | $S_{.025}$ | $S_{.01}$ | Significance |
|---|---|---|---|---|---|---|
| $S_{static}$ | $S_{tracked}$ | 243.5 | 238 | 248 | 260 | Yes: 0.05 |
| $S_{static}$ | $S_{custom}$ | 288.5 | 238 | 248 | 260 | Yes: 0.01 |
| $S_{tracked}$ | $S_{custom}$ | 235 | 238 | 248 | 260 | No |
| $U_{static}$ | $U_{tracked}$ | 218.5 | 238 | 248 | 260 | No |
| $U_{static}$ | $U_{custom}$ | 263 | 238 | 248 | 260 | Yes: 0.01 |
| $U_{tracked}$ | $U_{custom}$ | 225.5 | 238 | 248 | 260 | No |
| $E_{static}$ | $E_{tracked}$ | 266 | 238 | 248 | 260 | Yes: 0.01 |
| $E_{static}$ | $E_{custom}$ | 301 | 238 | 248 | 260 | Yes: 0.01 |
| $E_{tracked}$ | $E_{custom}$ | 213 | 238 | 248 | 260 | No |

The results from Table 1 indicate that static and custom were statistically significant, and that custom was more liked by participants than static. Tracked has both a significant increase in satisfaction and ease of viewing, when compared to static. Custom scored better than tracked, while tracked and custom were not statistically significant.

The static and tracked implementations had one open-ended question, which asked for general feedback and any comments. The custom implementation had several open-ended questions. For example, some asked respondents to explain their answers to the Likert scale questions for each feature (change position, transparency, and resizing). The responses seemed to support the fact that the static interpreting implementation was the least desirable one of the three, with multiple complaints about the interpreter video being too far from the lecture video and about missing information because of inability to focus on two areas of interests at once. The following participant quotes capture this phenomenon:

"One problem is when I looked away from interpreter video to read math, I missed what interpreter said."

"I ignored the interpreting cuz I just looked at the content."

People liked the tracked implementation better than the static one, but the comments showed space for improvement. Several people said that the interpreter window moved too quickly and that the movement should be more gradual. Some people still had a hard time looking at both the interpreter and the lecture even with a closer interpreter:

"I like interpreter video to tracked a professor's list. One problem is when I looked away from interpreter video to read math, I missed what interpreter said."

"It would be nice if the interpreter video is not limited to the left side of the lecture video."

The customizable implementation had more positive response than the other two implementations. People liked the ability to change features and the control over the interpreter video:

"I liked that i could adjust the size and make it transparent so the interpreter can be right on top of the notes. I think it would also be helpful if the interpreter background matched the color of the notes."

"It is very easy and simple to move the interpreter with mouse."

## 5. Conclusions

We found that many people did indeed like the tracked and customizable closed interpreting implementation over the static one. While there were several survey results and feedback that disagreed with each other, in general, there was a noticeable and significant increase in satisfaction, understanding, and ease of viewing when comparing the static implementation to the tracked implementation, and when comparing the tracked implementation to the customizable implementation. Participants also preferred the interpreting features that allow for resizing and relocating the interpreter video in the customizable implementation, but the transparency feature was not as well-received compared to the other two. Our study indicates that people preferred the tracked interpreting over the static interpreting, and the customizable interpreting over both the tracked and static interpreting views.

## 6. Future Work

Tracked video could be done automatically, by programming it to locate the most optimal place for interpreter. Other possibilities include addressing open-ended feedback, such as the background of the interpreter and the lecture video contrasting too sharply in the customizable implementation.

## 7. ACKNOWLEDGMENTS

This work was supported by the National Science Foundation Award IIS-1460894.